\documentclass[aps,pre,showpacs,showkeys,twocolumn,groupedaddress,floatfix]{revtex4-1}
\usepackage{graphicx}
\usepackage{amsthm,amsmath,amssymb}  % AMS
\usepackage{hyperref}
\begin{document}

\title{Percolation and jamming of linear $k$-mers on square lattice with defects:\\ effect of anisotropy}

\author{Yuri~Yu.~Tarasevich}
\email[Correspondence author: ]{tarasevich@asu.edu.ru}
\affiliation{Astrakhan State University, 20a Tatishchev Street, Astrakhan, 414056, Russia}

\author{Valeri~V.~Laptev}
\email{laptev@ilabsltd.com}
\affiliation{Astrakhan State University, 20a Tatishchev Street, Astrakhan, 414056, Russia}
\affiliation{Astrakhan State Technical University, 16 Tatishchev Street, Astrakhan, 414025, Russia}

\author{Andrei~S.~Burmistrov}
\email{ksairen10@mail.ru}
\affiliation{Astrakhan State University, 20a Tatishchev Street, Astrakhan, 414056, Russia}

\author{Taisiya~S.~Shinyaeva}
\email{tae4ka19@mail.ru}
\affiliation{Astrakhan State University, 20a Tatishchev Street, Astrakhan, 414056, Russia}

\author{Nikolai~V.~Vygornitskii}
\email{vygor@ukr.net}
\affiliation{Institute of Biocolloidal Chemistry named after F.D. Ovcharenko, NAS of Ukraine, 42 Boulevard Vernadskogo, 03142 Kiev, Ukraine}

\author{Nikolai~I.~Lebovka}
\email[Correspondence author: ]{lebovka@gmail.com}
\affiliation{Institute of Biocolloidal Chemistry named after F.D. Ovcharenko, NAS of Ukraine, 42 Boulevard Vernadskogo, 03142 Kiev, Ukraine}

\begin{abstract}
Using the Monte Carlo simulation, we study the percolation and jamming of oriented linear $k$-mers on a square lattice that contains defects. The point defects with a concentration, $d$, are placed randomly and uniformly on the substrate before deposition of the $k$-mers. The general case of unequal probabilities for orientation of depositing of $k$-mers along different directions of the lattice is analyzed. Two different relaxation models of deposition that preserve the predetermined order parameter $s$ are used. In the relaxation random sequential adsorption (RRSA) model, the deposition of $k$-mers is distributed over different sites on the substrate. In the single cluster relaxation (RSC) model, the single cluster grows by the random accumulation of $k$-mers on the boundary of the cluster (Eden-like model). For both models, a  suppression of growth of the infinite (percolation) cluster at some critical concentration of defects $d_c$ is observed. In the zero defect lattices, the jamming concentration $p_j$ (RRSA model) and the density of single clusters $p_s$ (RSC model) decrease with increasing length $k$-mers and with a decrease in the order parameter.
For the RRSA model, the value of $d_c$ decreases for short $k$-mers ($k < 16$) as the value of $s$ increases. For $k=16$ and 32, the value of $d_c$ is almost independent of $s$. Moreover, for short $k$-mers, the percolation threshold is almost insensitive to the defect concentration for all values of $s$.
For the RSC model, the growth of clusters with ellipse-like shapes is observed for non-zero values of $s$.  The density of the clusters $p_s$ at the critical concentration of defects $d_c$ depends in a complex manner on the values of $s$ and $k$. An interesting finding for disordered systems ($s=0$) is that the value of $p_s$ tends towards zero in the limits of the very long $k$-mers, $k \to \infty$ and very small critical concentrations $d_c \to 0$. In this case, the introduction of defects results in a suppression of $k$-mer stacking and in the formation of `empty' or loose clusters with very low density. On the other hand, denser clusters are formed for ordered systems with $p_s\approx 0.065$ at $s=0.5$ and $p_s\approx 0.38$ at $s=1.0$.
\end{abstract}



\maketitle

\section{Introduction}\label{sec:introduction}
Deposition of large particles such as colloids, polymers or nanotubes on substrates can be considered and studied as random sequential adsorption (RSA)~\cite{Evans1993RMP}. During RSA, objects randomly deposit on
the substrate; this process is irreversible, and the newly adsorbing objects cannot overlap or pass through previously deposited ones. The substrate may be prepatterned (e.g., see~\cite{Cadilhe2007JPhysCM}) or include some impurities (defects) (e.g., see~\cite{Lee1997PRE}).
The adsorbed objects may be identical or present a mixture of objects of different sizes and shapes (e.g., see~\cite{Lee2000,Budinski-Petkovic2011}). Moreover, anisotropy can be introduced by postulating unequal probabilities for the deposition of elongated objects along different axes (e.g., see~\cite{Budinski-Petkovic2011,Oliveira1992PRA}). The anisotropy of deposition can reflect the influence of external fields, flows or anisotropy of the substrate. The adsorption of the elongated particles in the presence of external fields produces anisotropic layers (e.g., see~\cite{PagonabarragaPRL1995,Pagonabarraga1999PRE}).

Very often, during RSA, a substrate can be considered as a discrete space, e.g., a regular or irregular lattice, or as a tree. The simplest, but most commonly used instance of a discrete substrate is the square lattice. The objects selected to be adsorbed onto a square lattice are generally linear in shape. Such linear segments (also denoted as needles, linear chains, stiff rods or $k$-mers) consist of $k$ successive connected sites.

If deposition of the objects can continue indefinitely, a jamming state is reached. At the jamming state, there are still voids on the substrate between the previously deposited objects, but their sizes or shapes are not sufficient to allow the deposition of even one additional object.

For a perfect lattice (a lattice without any defects), the jamming coverage, $p_j$, is the fraction of the sites occupied by deposited objects. For a diluted (disordered, disturbed) lattice (a lattice containing defects or impurities), there are several ways to define the jamming coverage~\cite{Ben-Naim1994JPhysA,Lee1996JPhysA,Cornette2006JCP}.
In the present research, we have used only the pure object jamming limit and denoted it as the jamming concentration, $p_j$, for short. The pure object jamming limit, $p_j $, is defined as the fraction of the total number of lattice sites occupied by the deposited objects~\cite{Lee1996JPhysA}.

Ben-Naim and Krapivsky obtained an important analytical result for the jamming concentration in a one-dimensional case~\cite{Ben-Naim1994JPhysA}.

If the concentration of the deposited objects on the substrate is sufficiently large, there may be a path through the objects from one side of the system to its opposite side. Below this concentration, there is no spanning
path through the system, while above this concentration, there is a connected component of the order of the size of the system. This concentration is called the percolation threshold. (See, e.g.,~\cite{Stauffer} for the details; art of the state may be found, e.g., in~\cite{Araujo2014EPJST}.) For some systems,  percolation never occurs, even at the jamming concentration (e.g., when the adsorbed layer is produced by the deposition of equally-sized squares on a square lattice and where the side of the square is greater than the length of 3 sites~\cite{Nakamura1987PRA}). In fact, the percolation threshold corresponds to a phase transition, e.g., insulator-conductor.  Different definitions of the percolation threshold are used for lattices with defects similar to the definitions of the jamming concentration. In the present research, we have used the term `percolation threshold' ($p_c$) as  the ratio of the sites occupied by the objects to the total  number of lattice sites.

Cornette et al.~\cite{Cornette2006JCP,Cornette2011PhysA} investigated numerically the percolation of polyatomic species in the presence of impurities on a square lattice with periodic boundary conditions. Bond and site percolation problems have been taken in consideration. Linear $k$-mers as well as so-called SAW $k$-mers, i.e. segments of self avoiding walk, have been studied up to $k=9$. A phase diagram where the critical concentration of impurities is plotted as a function of $k$ has been offered.

The kinetics of the random sequential adsorption of line $k$-mers (with values of $k$ up to 64) has been studied on a disordered substrate occupied by the point impurities~\cite{Lee1996JPhysA}. The coverage of the surface and the jamming limits are calculated by the Monte Carlo method. The coverage has an asymptotically exponential behavior at low concentration of the impurities. The jamming limits depend on the concentrations of the impurities, $d$. At $d < d^*$ the jamming limits decrease as the value of $d$ increases. At $d>d^*$ the jamming limits increase as the value of $d$ increases, where the value of $d^*$ depends on $k$. In the one-dimensional case, the results are in good agreement with the published analytical results~\cite{Ben-Naim1994JPhysA}. The coverage and the jamming limits on a two-dimensional disordered lattice are similar to the one-dimensional case. The jamming limits decrease monotonically as the length of the line segments increases. %The minima locations of the jamming limit for both one and two dimensions are at the same values for a given length of the $k$-mer.

Recently, the impact of defects on percolation in the random sequential adsorption of linear $k$-mers on square lattices was investigated for a particularly wide range of $k$-mers ($k$-mer lengths from 2 to 256)~\cite{Tarasevich2015PRE}. Two different cases were studied: (1) where it was assumed that the initial square lattice was nonideal and some fraction, $d$, of the sites was occupied by nonconducting point defects (impurities); (2) where it was assumed that some fraction, $d$, of the sites in the $k$-mers themselves consisted of defects, i.e., was nonconducting, while the initial square lattice was perfect.

Mixed site-bond percolation was studied for the RSA of $k$-mers on heterogeneous lattices with variable connectivity, $z$~\cite{Quintana2006}. The simulations were performed for $k=1$--$3$ on a triangular lattice. Percolation phase diagrams in terms of the percolation threshold $p_c$ versus lattice connectivity, $z$, were obtained. For the RSA deposition of monomers onto a triangular lattice with defects preliminarily filled with $k$-mers, the percolation ($k=3$--$24$)~\cite{Kondrat2005} and the jamming ($k\leq 50$)~\cite{Kondrat2006} were investigated by means of Monte Carlo simulations. The nonmonotonicity of the percolation threshold as a function of the impurity concentration was observed~\cite{Kondrat2005}.

The RSA of polydisperse mixtures of $k$-mers has also been extensively   investigated~\cite{Budinski-Petkovic2011,Dolz2007, Budinski-Petkovic2012}. A phase diagram separating percolating from non-percolating regions for mixtures of monomers and $k$-mers on a square lattice ($k=2$--$7$) has been also determined~\cite{Dolz2007}. The jamming coverage for a mixture  has been found to be greater than the jamming coverage either of the components making the mixture~\cite{Budinski-Petkovic2011}. On the other hand, the percolation threshold for a mixture was slightly greater than that of the longest $k$-mer~\cite{Budinski-Petkovic2012}. The continuum random sequential adsorption of polymer on a flat and homogeneous surface has also  been studied~\cite{Ciesla2013PRE}. The polymers were  modeled as stiff or flexible chains of monomers.

However, most of the previous studies have been devoted to the 2D deposition of randomly oriented anisometric particles on the substrates. In several works,  problems with unequal probabilities for the orientation of the deposition of $k$-mers along different directions of the lattice have also been analyzed. The degree of anisotropy can be characterized by the order parameter, $s$, defined as
\begin{equation}\label{eq:defs}
s = \frac{\left| N_| - N_- \right|}{N_| + N_-},
\end{equation}
where  $N_|$ and $N_-$ are the numbers of vertically and horizontally oriented particles.

The anisotropic RSA of dimers on a square lattice has additionally been investigated using both Monte Carlo simulation and  time series expansion~\cite{Oliveira1992PRA}.
The problem of the anisotropic RSA of dimers on a square lattice has been  studied~\cite{Cherkasova2010}. Data from the Monte Carlo simulations evidence that the  orientational order parameter, $s$, influences both the values of the percolation threshold $p_c$ and the jamming concentration $p_j$. In particular, the value of  $p_j$ decreases as the order parameter, $s$ , increases~\cite{Oliveira1992PRA,Cherkasova2010}. An interesting finding is that in the limit of $s\to 1$, the asymptotic fraction of dimers with horizontal direction does not vanish but equals $e^{-2}[1-\exp(-2e^{-2})]/2\approx 0.016\,046$~\cite{Oliveira1992PRA}. The properties of the anisotropic RSA of flexible $k$-mers on a 2D triangular lattice have been studied numerically by means of Monte Carlo simulations~\cite{Budinski-Petkovic2011}. It was shown that the relaxation time to the jamming limit increases with the degree of anisotropy of the elongated particles.

The effect of anisotropic deposition of $k$-mers on the jamming~\cite{Lebovka2011PRE} and the percolation~\cite{Tarasevich2012PRE} has also been intensively studied by means of Monte Carlo simulations.
These detailed studies revealed that the RSA model does not allow preservation of the order parameter, $s$, and, in fact, the substrate `selects' the $k$-mer with appropriate orientation, resulting in deviation of the predetermined order parameter, $s$, from the one actually observed one, $s_0$. A special variant of relaxation for the RSA model (the RRSA model) with better preservation of the predetermined anisotropy has been developed~\cite{Lebovka2011PRE}. In the RRSA model, the binding of a $k$-mer to the adsorbing substrate is strong, and the $k$-mer has additional possibilities for joining the surface after an unsuccessful attempt.

Our aim was to study the influence of anisotropy on the percolation threshold and the jamming concentration of linear $k$-mers onto a square lattice with point defects. The results for the multiple cluster relaxation random sequential adsorption (RRSA) model and the single cluster relaxation (RSC) model were compared. In the RRSA model, the deposition was distributed over different sites of the lattice, whereas in the RSC model the cluster grows at its perimeter, starting from an active seed in the center of the lattice. In fact, the RSC  model is an extension of the Eden growth model~\cite{Eden1961} for the case of $k$-mers on a lattice with defects.

The rest of the paper is organized as follows. Section~\ref{sec:method} describes our RRSA and RSC models of $k$-mer deposition. The results obtained using finite size scaling theory and dependencies of the percolation threshold, $p_c$, and the jamming concentration, $p_j$, vs the order parameter, $s$, and the defect concentration, $d$, are examined and discussed in detail in Section~\ref{sec:results}. We summarize the results and conclude the paper in Section~\ref{sec:discussion}.
%\clearpage

\section{Method}\label{sec:method}

In both the RRSA and the RSC models, the square lattices were initially randomly filled with point defects at a given concentration, $d$. Before deposition of the $k$-mers, we choose an appropriate orientation in the horizontal or vertical directions according to the given value of the order parameter $s$. The lattice sites were then occupied by the addition of $k$-mers. The defects and previously deposited $k$-mers can inhibit the deposition of newcomers. If the attempt was unsuccessful, a new lattice site (the RRSA model) or a new empty cluster perimeter site (the RSC model) was randomly selected until the object could be deposited. The objects may move over all the substrate (RRSA model) or over all the cluster perimeters (RSC model) searching for a sufficiently large empty space. The other specific details are presented below.

\subsection{RRSA model}\label{subsec:RSA}

In the RRSA model, the deposition terminates when a jamming state is reached along one direction~\cite{Lebovka2011PRE}. We considered a lattice with periodic (toroidal) boundary conditions to eliminate the border effects and, in contrast to~\cite{Newman2001PRE}, treated spiral clusters as wrapping (percolating). We checked the percolation in two perpendicular directions and used two criteria: there is percolation in both directions (criterion AND), there is percolation at least along one direction (OR).
\begin{figure*}
  \centering
  \includegraphics[width=\textwidth,keepaspectratio]{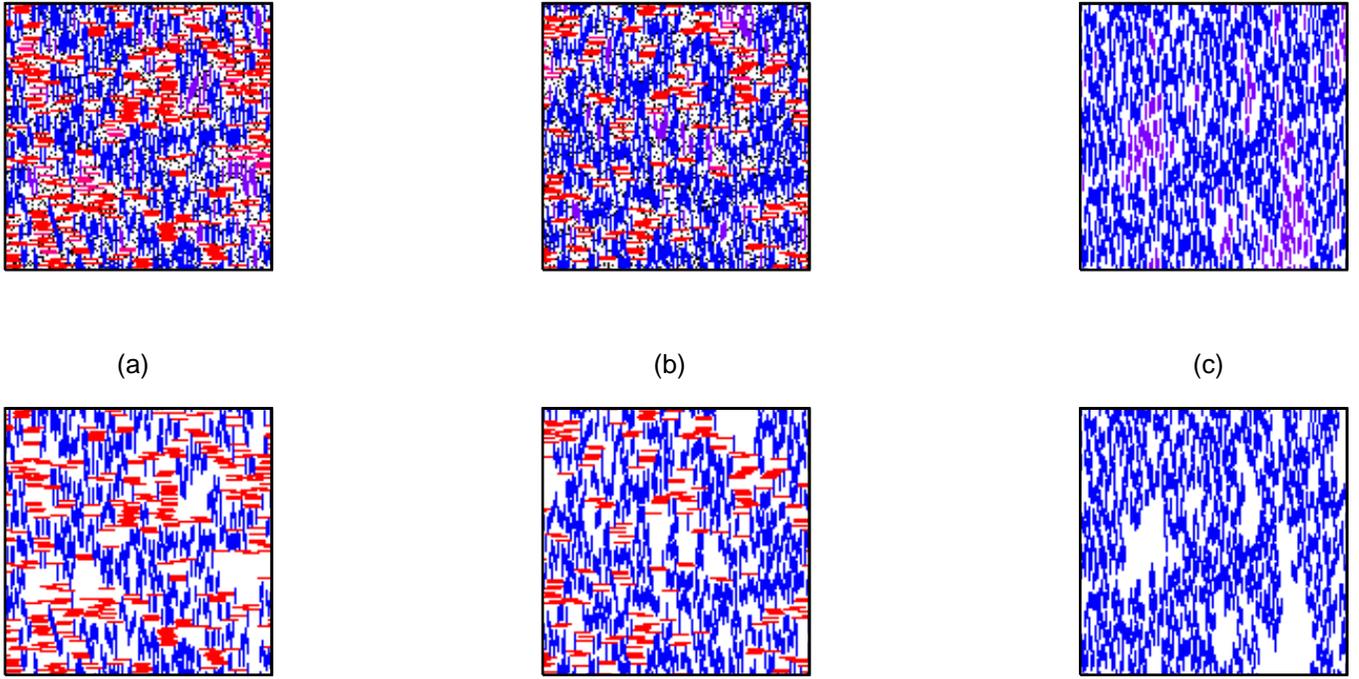}
  \caption{Examples of the jamming states (upper row) and the percolation clusters (bottom row) of anisotropically deposited $k$-mers on a square lattice with defects. (a) $s=0.25$, (b) $s=0.5$, (c) $s=1$. The lattice
size is $1024\times1024$, $k = 8$, a fragment of the lattices with $128\times128$ sites is shown. The concentration of defects on the lattice is
0.09. Online: Horizontal $k$-mers are shown in red, vertical $k$-mers are shown
in blue, $k$-mers not belonging to the percolation cluster are shown in the same colors but with a different hue, empty sites are
shown in white, defects on the lattice are shown in black.
Print: gray-scale.}\label{ fig:percjam }
\end{figure*}

Examples of the jamming states (upper row) and the percolation clusters (bottom row)  at different values of the order parameter, $s$, are presented in  Fig.~\ref{ fig:percjam }. The concentration, $d$, of defects on the lattice is 0.09.

\begin{figure*}[htbp]
  \centering
  (a)\includegraphics[width=0.45\linewidth]{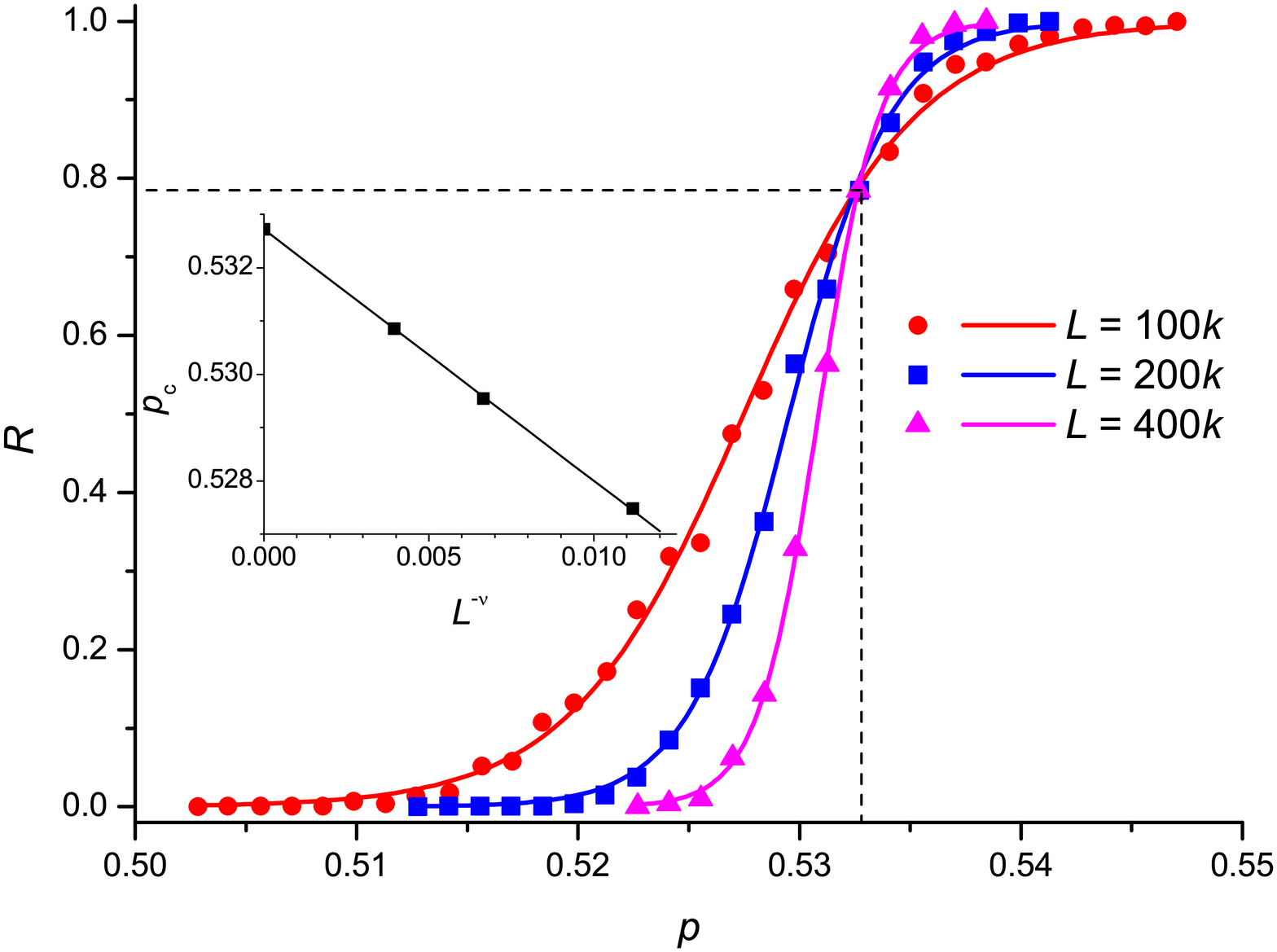} (b)\includegraphics[width=0.45\linewidth]{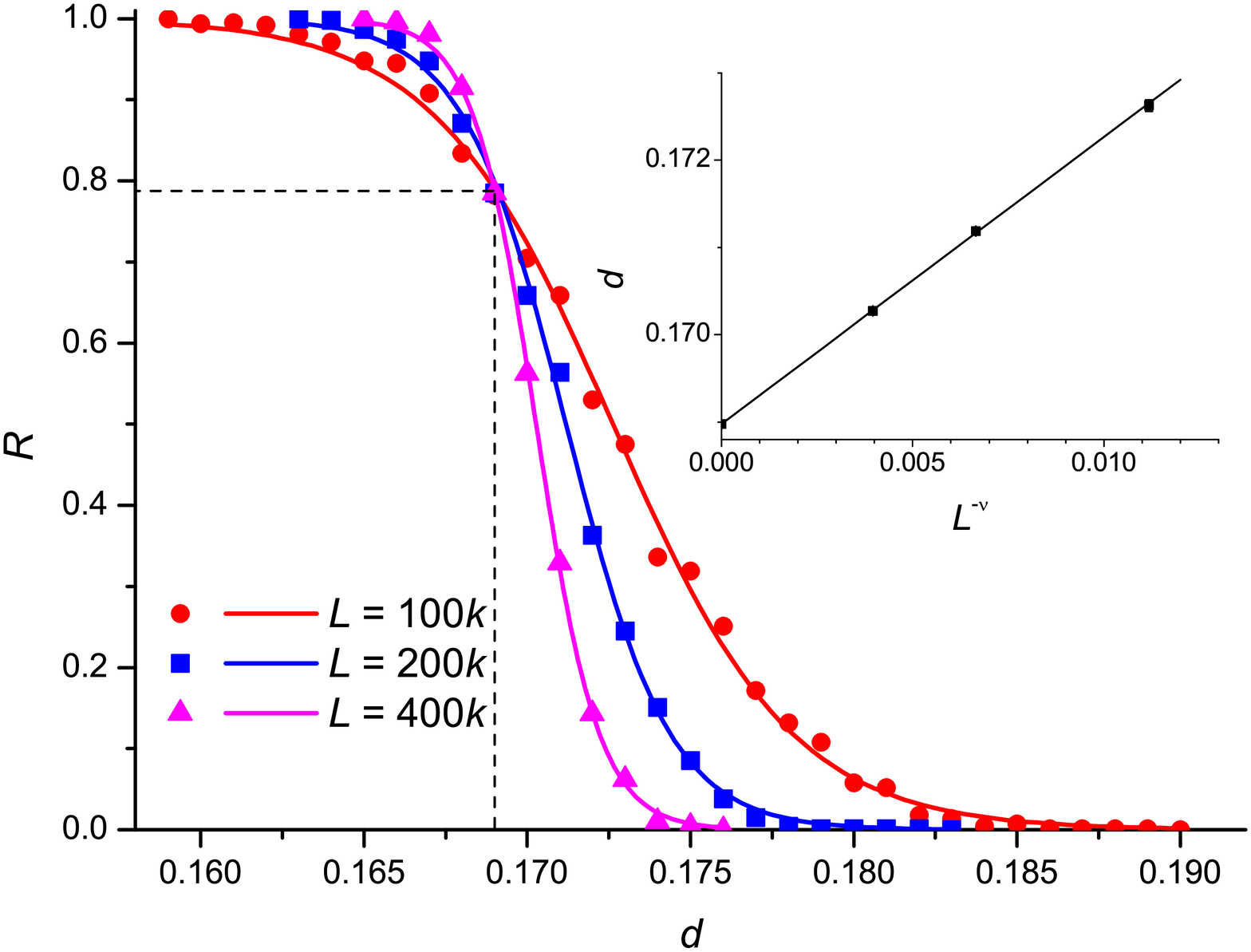}\\
  \caption{Probability curves for $k=4$, $s=0.75$. (a) $R(p)$. (b) $R(d)$. Insets: scaling. The statistical error is smaller than the marker size.}\label{fig:scaling}
\end{figure*}

For each given order parameter, $s$, and concentration of defects, $d$, we filled the lattice with $k$-mers to a concentration $p$, or to  jamming and checked whether there was percolation. We repeated this 1000 times and found the probability, $R(p)$, that percolation occurs at a particular concentration of the $k$-mers. The abscissa of the inflection point of the curve was treated as the estimate of the percolation threshold for the given lattice size. We used  lattices of  sizes $L = 100k, 200k, 400k$ and performed finite-size analysis to obtain the percolation threshold at the thermodynamic limit ($L \to \infty$) $ p_c \propto L^{-1/\nu}$, where $\nu = 4/3$ is the critical exponent of the correlation length for the 2d random percolation problem~\cite{Stauffer}. All probability curves intersect each other at one point with the co-ordinates $(p_c,R^*)$, where $R^* = R(p_c)\approx 0.78$. The value of $R^*$ depends on how the percolation is defined~\cite{Newman2001PRE}. The intersection point offers another way to estimate the percolation threshold. The examples of probability curves and scaling are shown in Fig.~\ref{fig:scaling}.

We compared our numerical simulation for completely aligned $k$-mers ($s=1$) with the analytical results~\cite{Ben-Naim1994JPhysA} and found that there is not  visible difference between the analytical and numerical results.

\subsection{RSC model}\label{subsec:AL}

The Eden algorithm~\cite{Eden1961}
was applied to grow a single cluster from an active seed $k$-mer on the square lattice. In the RSC model, the initial seed $k$-mer is placed at the center of the lattice and it has $2k+2$ initial perimeter sites.  The RSC algorithm uses the following steps:
\begin{description}
  \item[Step 1] Randomly choose one empty cluster perimeter site and try to add a new $k$-mer to any available lattice sites, if such the deposition is not inhibited. Repeat this step using the same orientation of $k$-mer until successful attempt.
  \item[Step 2] Determine the new cluster perimeter sites.
  \item[Step 3] Continue the previous steps until there remain no untested cluster perimeter sites or the cluster reaches one of the four boundaries of the lattice.
\end{description}

\begin{figure*}
  \centering
  \includegraphics[width=\textwidth]{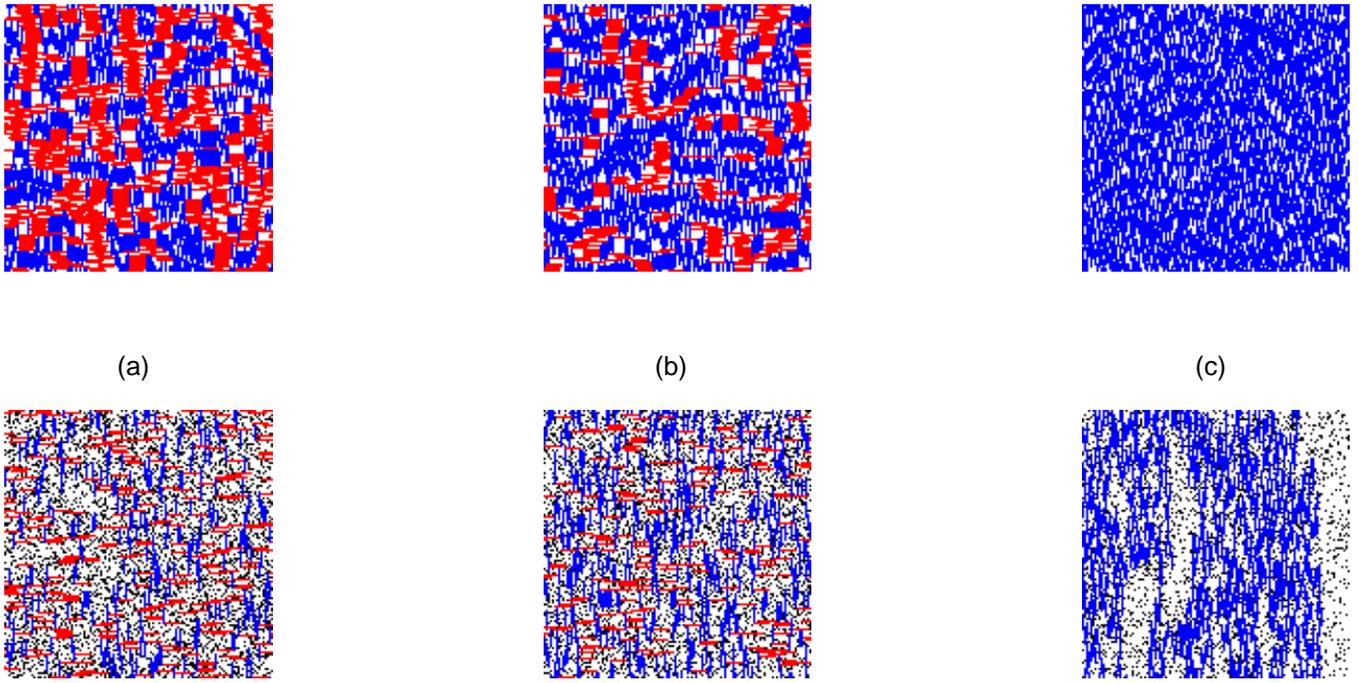}\\
  \caption{Examples of the clusters in the absence (upper row) and in the presence (bottom row) of defects for the RSC model at different values of the order
parameter: (a) $s=0.0$, (b) $s=0.5$, (c) $s=1$. The lattice size is $1024\times1024$, $k = 8$, a fragment of the lattices with $128\times128$ sites is shown. The concentration of defects on the lattice, $d$, is near the critical $d_c$: 0.215 ($s=0.0$), 0.186 ($s=0.5$) and 0.134 ($s=1.0$).
Online: Horizontal $k$-mers are shown in red, vertical $k$-mers are shown in blue, empty sites are
shown in white, defects on the lattice are shown in black. Print: gray-scale.}\label{fig:f01}
\end{figure*}

\begin{figure*}%[htbp]
  \centering
    (a)\includegraphics[width=0.45\linewidth]{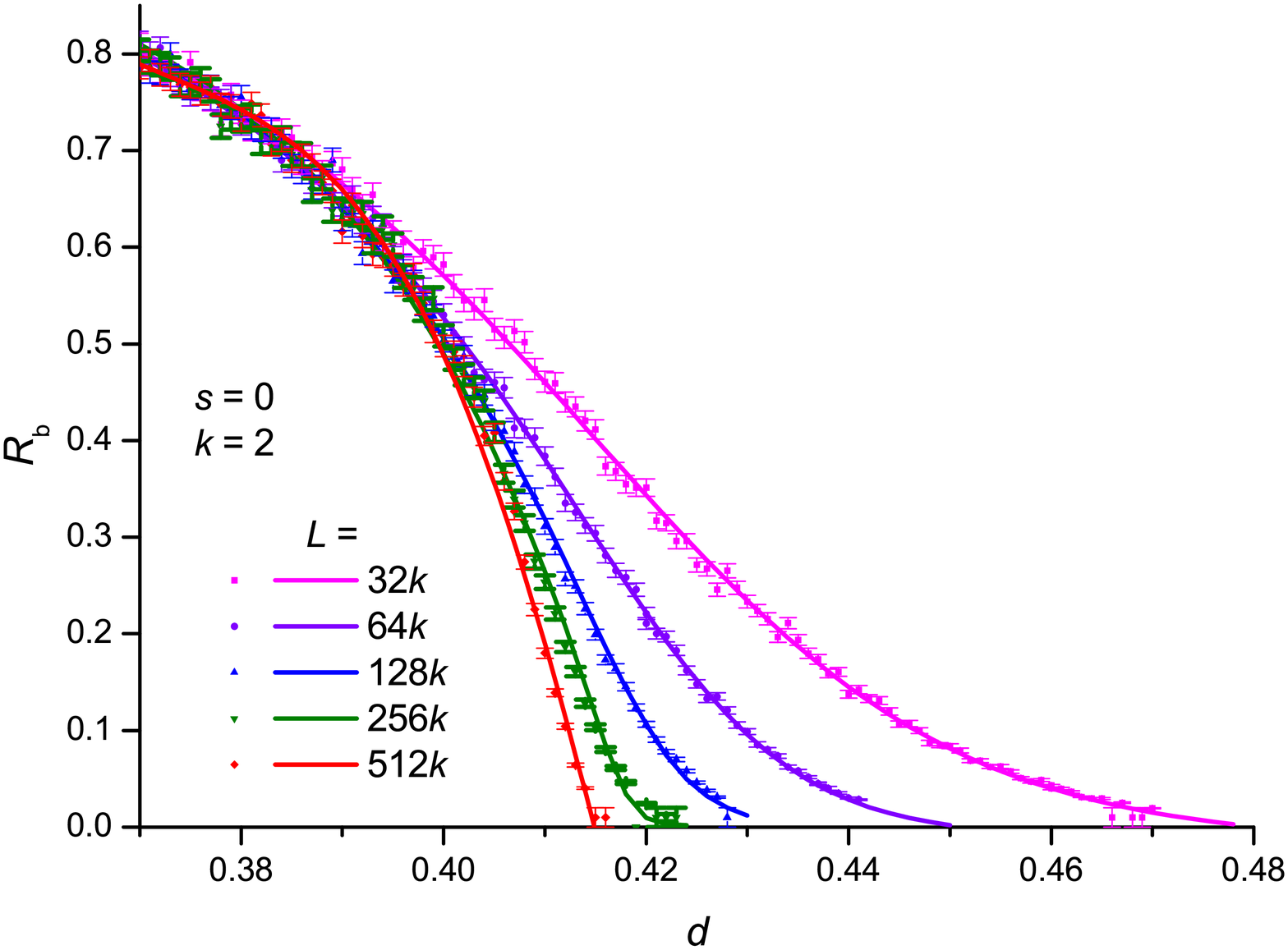} (b)\includegraphics[width=0.45\linewidth]{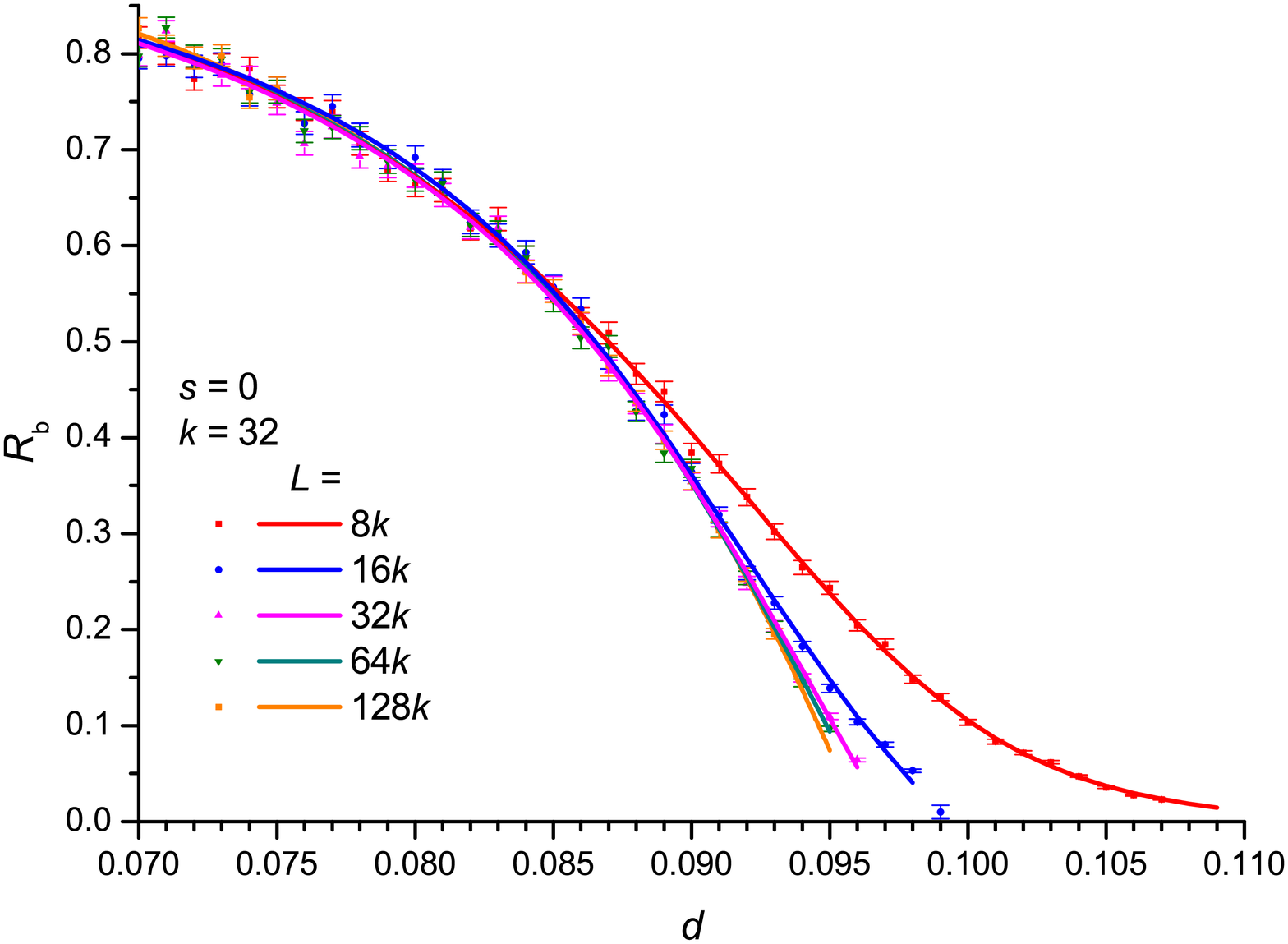}\\
  \caption{Probability curves $R_b(d)$ for the RSC model, $k = 2$ (a) and $k = 32$ (b). The data are presented for disordered systems, $s = 0$. The statistical errors are smaller than the marker size. The lines are provided simply as eye-guides. }\label{fig:f02}
\end{figure*}

Examples of clusters in the absence and presence of defects for the RSC model ($k=8$) with different order parameters, $s$, are shown in Fig.~\ref{fig:f01}. When defects are absent, an infinite cluster can grow on the lattice and the formation of stacks of identically oriented $k$-mers (horizontal and vertical) can be observed. These stacks are rather similar to that observed for the RSA or the RRSA models~\cite{Lebovka2011PRE,Tarasevich2012PRE}. Visual observations of the patterns at different concentrations of defects allow us to draw the conclusion that the presence of defects restricts the formation of stacks and can prohibit the growth of an infinite single cluster for some critical concentration of defects, $d_c$.

\begin{figure*}[htbp]
  \centering
    (a)\includegraphics[width=0.45\linewidth]{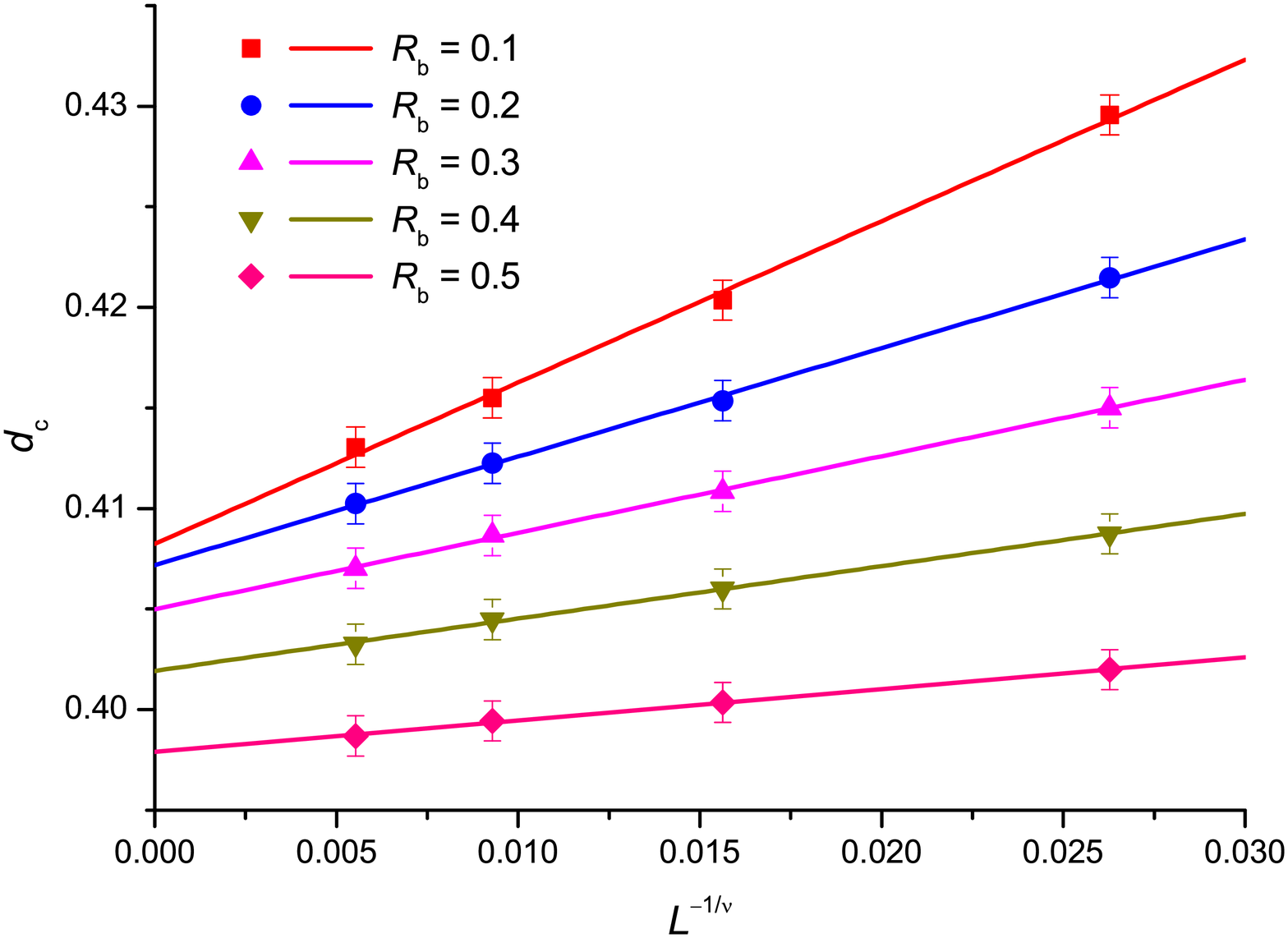} (b)\includegraphics[width=0.45\linewidth]{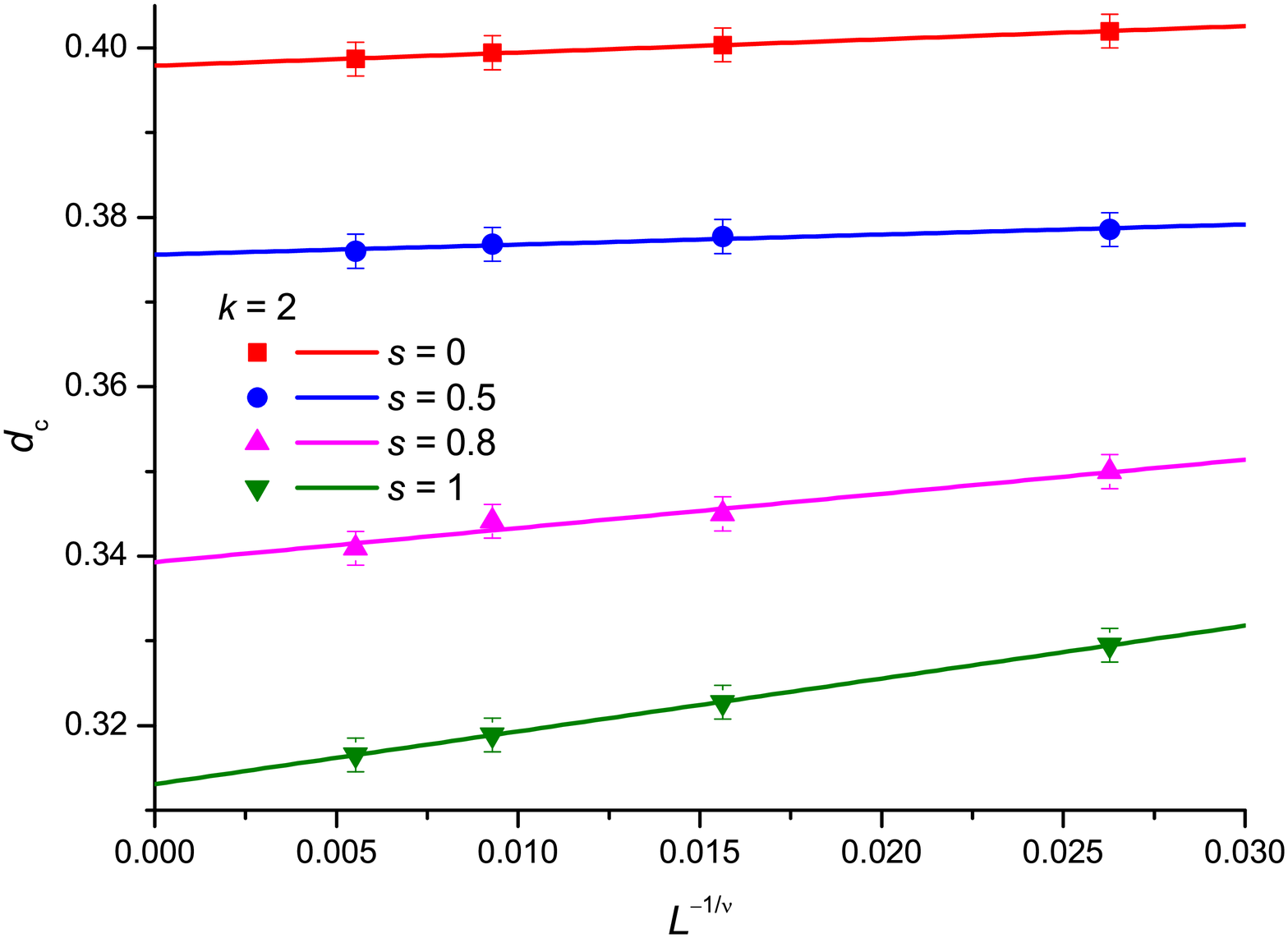}\\
  \caption{Concentration of defects, $d_c$, versus the scaled size of the system $L^{1/\nu}$ for the RSC model ($k=2$) at different values of probability $R_b$ ($s=0$) (a) and at different values of the order parameter $s$ ($R_b=0.5$) (b). Here, $\nu= 4/3$  is the critical exponent of the correlation length~\cite{Stauffer}.
}\label{fig:f03}
\end{figure*}

For each given anisotropy, $s$, and  concentration of  defects, $d$, we repeated the Monte Carlo experiments 1000 times and found the probability, $R_b(d)$, that the single cluster reaches one of the boundaries of the lattice at a given concentration of defects, $d$. We used lattices of different sizes in the interval $L = 8k$--$2048k$. Examples of the probability $R_b(d)$ curves for a disordered system ($s=0$) using two different values of $k$ are shown in Fig.~\ref{fig:f02}. The values of $R_b$ decrease as the concentration of defects $d$ increases. The commonly applied procedure for determination of the critical concentration $d_c$ for blocking of the cluster growth is to use finite scaling analysis for a fixed value of $R_b$ (e.g., at $R_b=0.5$). However, for the RSC model the probability $R_b(d)$ curve is not step-like even for infinite systems in the thermodynamic limit $L \to \infty$. For the RSC model at $L \to \infty$, $R_b(d)$ the curve smoothly descends as the value of $d$ increases reflecting the finite probability of the blocking of cluster growth from the initial seed $k$-mers surrounded by lattice defects. Examples of finite-size analysis at different values of $R_b$ ($k=2$, $s=0$) are presented in Fig.~\ref{fig:f03}a. Good linear dependencies of data were always observed in the $d$ versus  $L^{-1/\nu}$ coordinates, where $\nu = 4/3$ is the critical exponent of the correlation length for the 2d random percolation problem~\cite{Stauffer}. To be definitive, we always estimate the critical value of the concentration of defects $d_c$  in the limit $L \to \infty$ at the fixed value $R_b=0.5$ (see Fig.~\ref{fig:f03}a). At this value $d_c$, the growth of the infinite cluster is suppressed with a probability of 50\%. Examples of the $d (L^{1/\nu})$ dependencies for the RSC model ($k=2$) at different values of the order parameter $s$ ($R_b=0.5$) are presented in Fig.~\ref{fig:f03}b. The data shows that the slope of the scaling can be greatly dependent on the value of $s$.

The mean degree of the cluster anisotropy, $\delta$, has been calculated as
\begin{equation}
\delta =(r_y-r_x)/r,
\label{eq:delta}
\end{equation}
where  $r_y$ and $r_x$ are the radii of gyration of the cluster in $y$ and $x$ directions, respectively, and $r$ is the mean radius of gyration.

The density of the cluster of $k$-mers, $p_s$, was calculated inside a rectangle of size $2r_y$ and $2r_x$ with its center located at the geometrical center of the cluster. A finite scaling analysis was also carried out for the estimation of $p_s$ in the limit $L \to \infty$.

\section{Results}\label{sec:results}
\subsection{RRSA model}\label{subsec:RSAresults}

\begin{figure}[htbp]
  \centering
 \includegraphics[width=0.9\linewidth]{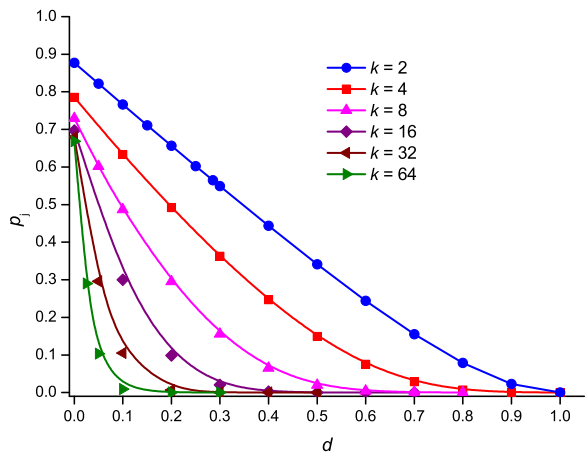}
   \caption{The jamming concentration, $p_j$,  versus the defect concentration, $d$, for different values of $k$ and a fixed order parameter $s=0.5$. The solid lines present are simply as an eye-guide. The statistical error is smaller than the marker size.} \label{fig:pjvsd}
\end{figure}

Figure~\ref{fig:pjvsd} shows the typical dependencies of jamming concentration, $p_j$,  versus the defect concentration, $d$, for different values of $k$ and a fixed value of the order parameter, $s=0.5$. For any given $k$, the value of $p_j$ decreases as the value of $d$ increases. The effect is more pronounced for the longer $k$-mers, and the jamming concentration decreases drastically with growth of $k$-mer length for any anisotropy. Similar dependencies were obtained for other  values of the order parameter, $s$.

\begin{figure}[htbp]
  \centering
  \includegraphics[width=0.9\linewidth]{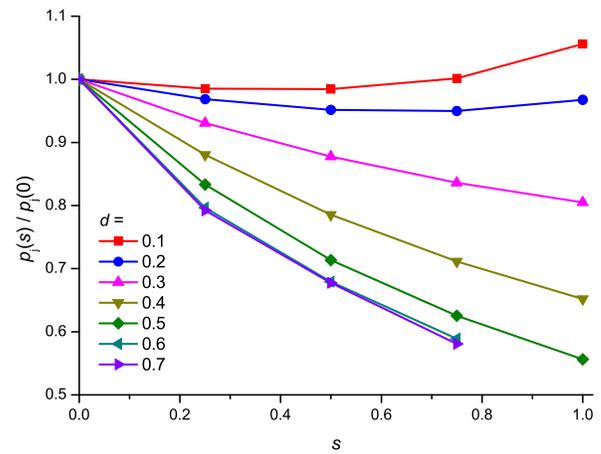}\\
  \caption{The relative jamming concentration, $p_j/p_j(s=0)$, versus the order parameter, $s$,at different concentrations of defects, $d$. The length of the $k$-mers equals 8. The solid lines are simply to provide eye-guidance. The statistical error is smaller than the marker size.} \label{fig:pjvssk=8}
\end{figure}

Figure~\ref{fig:pjvssk=8} shows the typical dependencies of the relative jamming concentration, $p_j/p_j(s=0)$, versus the order parameter, $s$, for different defect concentrations, $d$, for $k=8$. For large concentrations of defects ($d \ge 0.3$), the jamming concentration decreases monotonically as the extent of anisotropy increases, whereas  for small concentrations of defects ($d<0.3$), the jamming concentration has a minimum, near the value of $s = 0.5$.

\begin{figure}[htbp]
  \centering
  \includegraphics[width=0.9\linewidth]{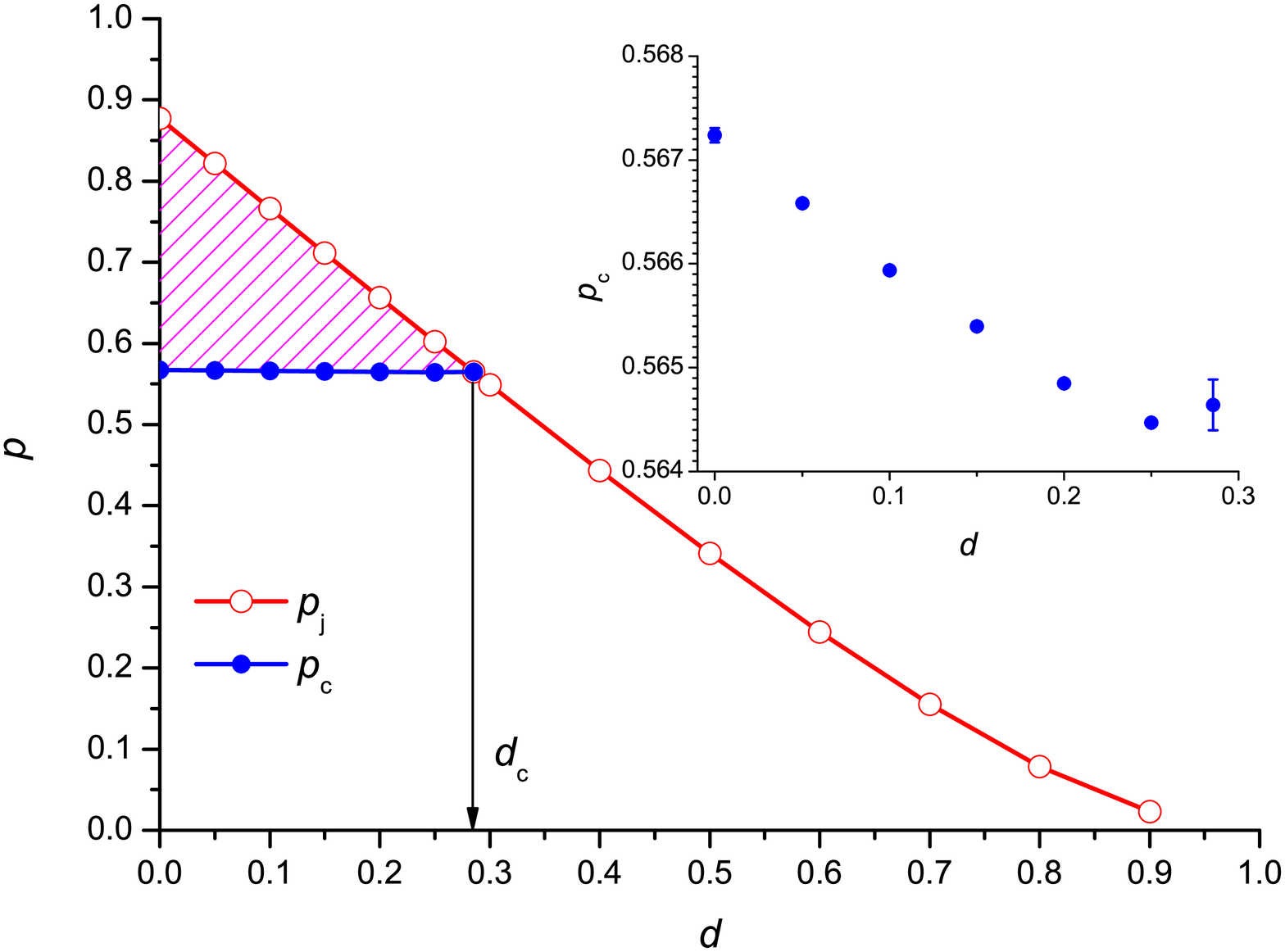}\\
  \caption{The percolation threshold, $p_c$, and the jamming concentration, $p_j$, as a function of the defect concentration, $d$, for the anisotropy $s=0.5$ and $k=2$. The hatched region corresponds to the concentrations of the objects and the defects when percolation is possible. Here $d_c$ is a critical concentration of defects that suppresses the formation of an infinite (percolation) cluster. The solid lines are simply present as eye-guides. The statistical error is smaller than the marker size when not shown explicitly.} \label{fig:k2pcpjvss}
\end{figure}

In the RRSA model, there exists a critical concentration of defects, $d_c$, above which the percolation, i.e. formation of an infinite cluster, is impossible even in the jammed state. An example of the dependencies of the percolation threshold, $p_c$, and jamming concentration, $p_j$, as functions of the defect concentration, $d$, for the given order parameter, $s=0.5$, and $k=2$ is presented in Fig.~\ref{fig:k2pcpjvss}.

For the short $k$-mers ($k\lessapprox 8$), the percolation threshold, $p_c$, is almost insensitive to the defect concentration, $d$, for any anisotropy, $s$~\cite{Tarasevich2015JPhCS} (e.g., see Fig.~\ref{fig:k2pcpjvss} for $k=2$). For long $k$-mers ($k>8$), $p_c(d)$ has a maximum between $d=0$ and $d=d_c$~\cite{Tarasevich2015PRE}.

\begin{figure}[htbp]
  \centering
  \includegraphics[width=0.9\linewidth]{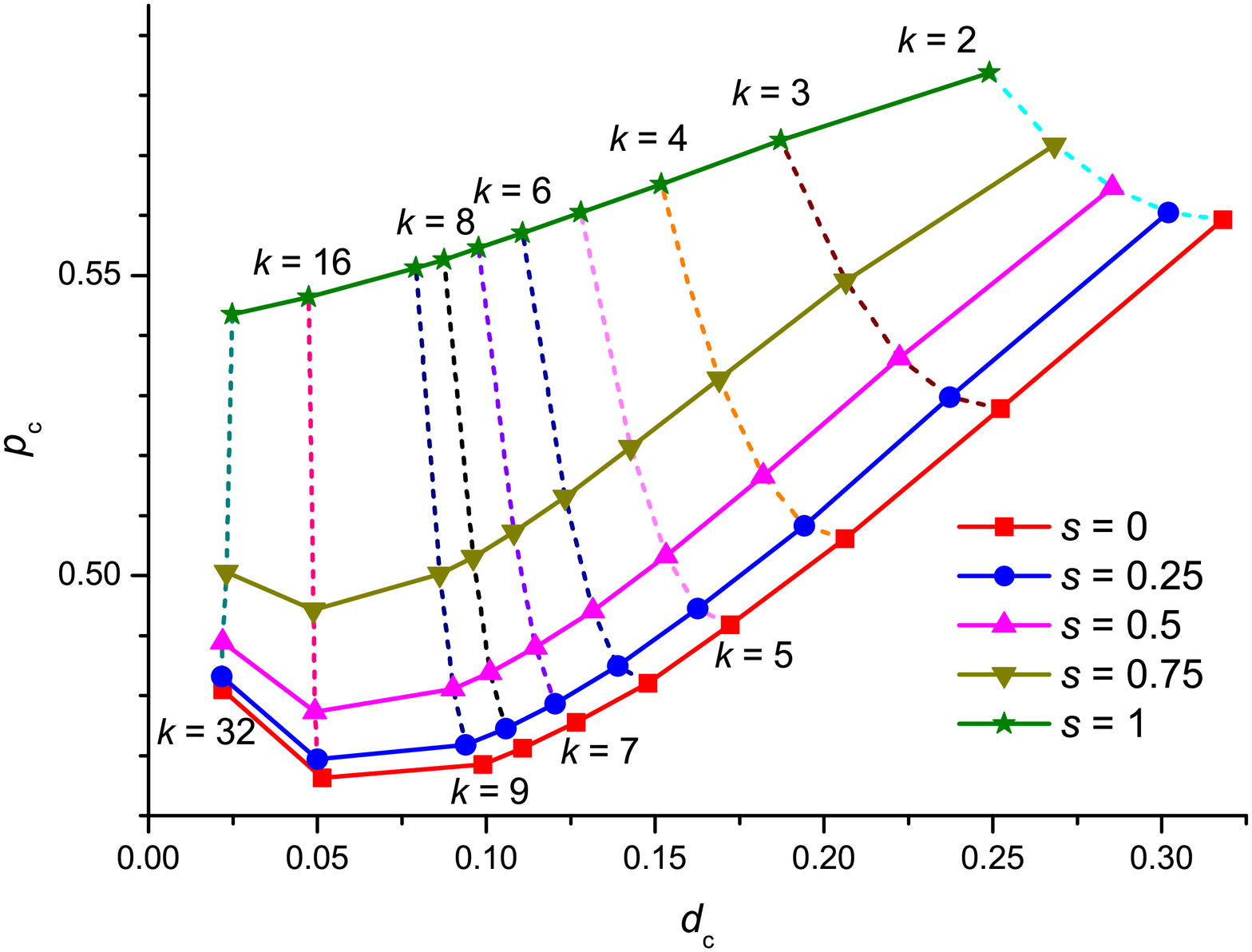}\\
  \caption{The percolation concentration, $p_c$, versus the critical concentration of defects, $d_c$, for different values of the order parameter, $s$, and lengths of the $k$-mers. The lines are simply present as eye-guides. The statistical error is smaller than the marker size.} \label{fig:dmvspm}
\end{figure}

Figure~\ref{fig:dmvspm} presents percolation phase diagram using as coordinates the critical percolation concentration, $p_c$, versus the critical concentration of defects, $d_c$, at different values of the order parameter, $s$, and lengths of the $k$-mers. The increase of order parameter, $s$, is always accompanied  by an increase of $p_c$. However, the $d_c$ versus $s$ behavior is different for short and long $k$-mers. Moreover, it is interesting that for partially oriented systems ($s<1$) the $p_c(d_c)$ dependencies at fixed values of $s$ go through a minimum at the values of $k$ lying in the interval between 9 and 16.

\subsection{RSC model}\label{subsec:RSCresults}
\begin{figure}[htbp]
  \centering
\includegraphics[width=0.9\linewidth]{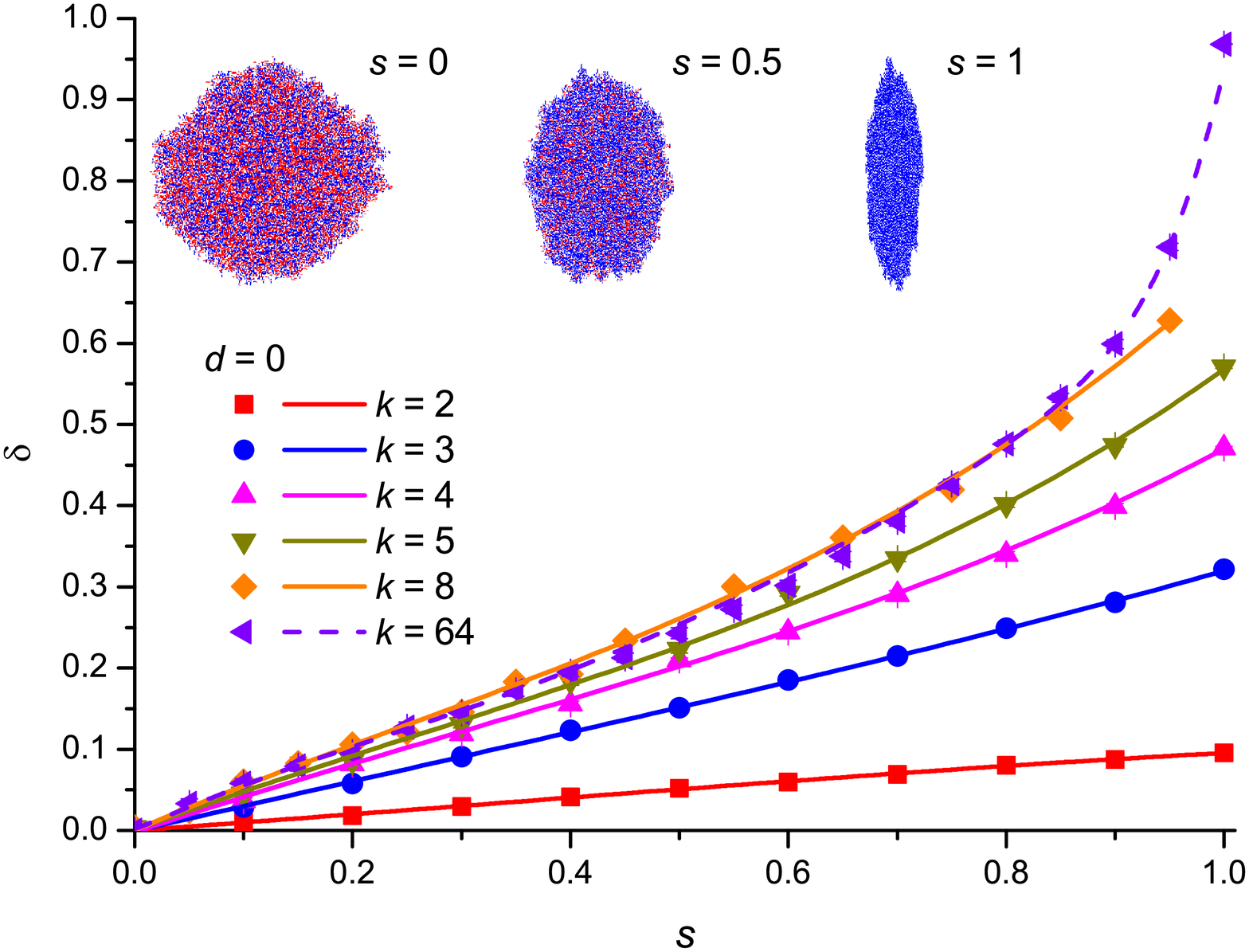}\\
\caption{Examples of the degree of the cluster anisotropy, $\delta$, versus the order parameter, $s$, for different values of $k$ for zero defect lattices ($d=0$). Inset: examples of the clusters for $k=8$.} \label{fig:f04}
\end{figure}

Preliminary investigations were performed for zero defect systems ($d=0$).
Figure~\ref{fig:f04} presents examples of the degree of the cluster anisotropy, $\delta$, versus the order parameter, $s$, for different values of $k$ for such zero defect systems ($d=0$). The value of $\delta$ increases as $s$ increases and the effects become more pronounced for the longer $k$-mers. At large values of $s$, the shape of the clusters becomes ellipse-like (see inset to Fig.~\ref{fig:f04}).

\begin{figure}[htbp]
  \centering
\includegraphics[width=0.9\linewidth]{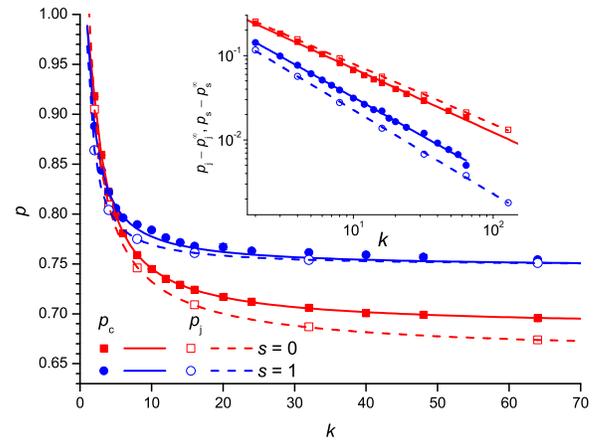}\\
\caption{Density of cluster, $p_s$, versus the value of $k$ for the RSC model, and the jamming concentration, $p_j$, versus the value of $k$ for the RRSA model~\cite{Lebovka2011PRE}. The data are presented for disordered $s=0$ and completely ordered $s=1$ zero defect systems ($d=0$). Inset: $p_s-p_s^\infty$  and $p_j-p_j^\infty$ versus $k$. Here, $p_s^\infty$  and $p_j^\infty$ are the limiting values at $k\to \infty$.} \label{fig:f05}
\end{figure}

Figure~\ref{fig:f05} compares both the density of the cluster, $p_s$, versus $k$ for the RSC model and the jamming concentration, $p_j$,~\cite{Lebovka2011PRE} versus $k$ for the RRSA model. It is worth noting that the values of $p_s$ and $p_j$ decrease as an inverse power of the linear segment length $k$ (see, inset to Fig.~\ref{fig:f05}) and approach their limiting values $p_s^\infty$ and $p_j^\infty$ with increasing $k$:
\begin{equation}\label{eq:power}
p_{s,j}(k) = p_{s,j}^\infty+a/k^\alpha.
\end{equation}
The parameters $p_s^\infty$ and $p_j^\infty$, $a$ and $\alpha$ are presented in Table~\ref{tab:pjvsSkoo}.
\begin{table} % Table 1
\caption{\label{tab:pjvsSkoo}
Estimated parameters $p_s^\infty$ and $p_j^\infty$, $a$ and, $\alpha$ in the power law function Eq.~\ref{eq:power}.
The data are presented for RSC and the RRSA~\cite{Lebovka2011PRE} models at $s=0$ and $s=1$ and
for the one-dimensional (1D) model~\cite{Lebovka2011PRE}. In all cases the coefficient of determination, $\rho$, was greater than $0.999$.}
\begin{ruledtabular}
\begin{tabular}{c|cc|cc|cc}
&\multicolumn{2}{c|}{$p_{j}(k=\infty)$} & \multicolumn{2}{c|}{$a$} & \multicolumn{2}{c}{$\alpha$}\\
\hline
$s$  &RSC    & RRSA  & RSC   & RRSA  & RSC   & RRSA\\
\hline
0  &0.67(6)&0.652(8)& 0.41(8)&0.417(5)& 0.77(7)&0.713(7)\\
1&0.745(0)&0.747(2)&0.27(3)&0.235(3)&0.93(1)&1.016(6)\\
\hline
1d&\multicolumn{2}{c|}{0.74759792} &
\multicolumn{2}{c|}{0.227(7)} & \multicolumn{2}{c}{1.011(1)}\\
\end{tabular}
 \end{ruledtabular}
\end{table}
The exponent $\alpha$ is not universal and depends upon the order parameter, $s$. It had been suggested that that such power behavior indicates the presence of a fractal structure of the cluster networks in both the RSC and RRSA models with the fractal dimension $d_f = 2-\alpha$~\cite{Lebovka2011PRE}. For completely ordered systems ($s = 1$), the value of $d_f$ is close to $1$ as expected for the 1D problem. For disordered systems at $s=0$, the values of $\alpha$  are noticeably smaller than 1 for both the RSC and the RRSA
models and so the fractal dimension, $d_f$, in this case is higher than 1.
%\clearpage

\begin{figure}[htbp]
  \centering
\includegraphics[width=0.9\linewidth]{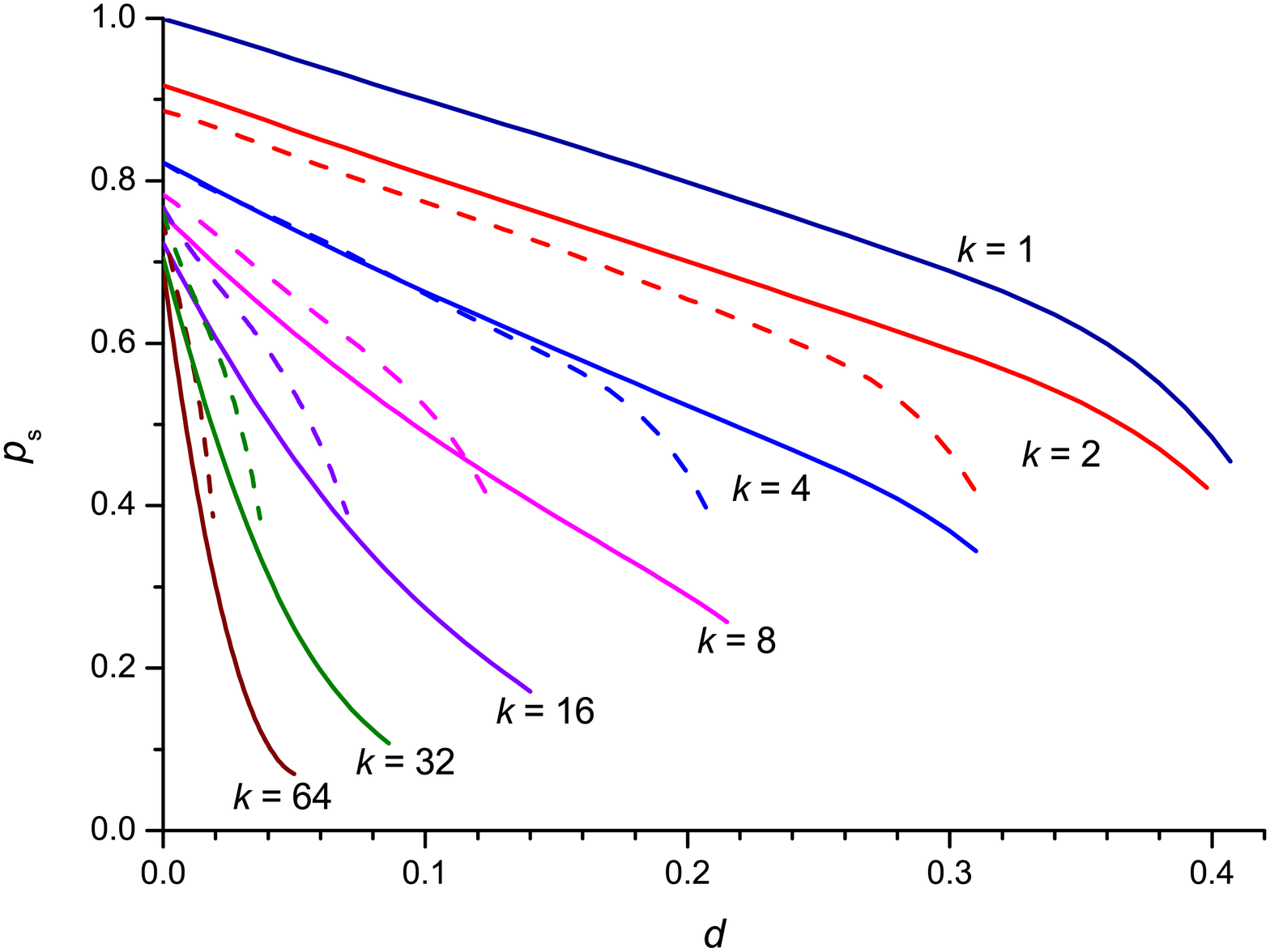}\\
\caption{Density of cluster, $p_c$, versus the concentration of defects, $d$, for the RSC model at different values of the linear segment length, $k$, for disordered $s=0$ (solid lines) and completely ordered $s=1$ (dashed lines) systems.} \label{fig:f06}
\end{figure}

 Figure~\ref{fig:f06} compares the density of the cluster, $p_s$, versus the concentration of defects, $d$, for the RSC model at different values of the linear segment length $k$  for disordered $s=0$ (solid lines) and completely ordered $s=1$ (dashed lines) systems. The values of $p_s$ decrease as the value of $d$ increases and there exists some limiting concentration of defects, $d_c$, that suppresses the growth of an infinite cluster. At any given value of $d$, the impact of ordering on the value of $p_s$ depends both upon the values of $k$ and $d$ (Fig.~\ref{fig:f06}).
\begin{figure}[htbp]
  \centering
\includegraphics[width=0.9\linewidth]{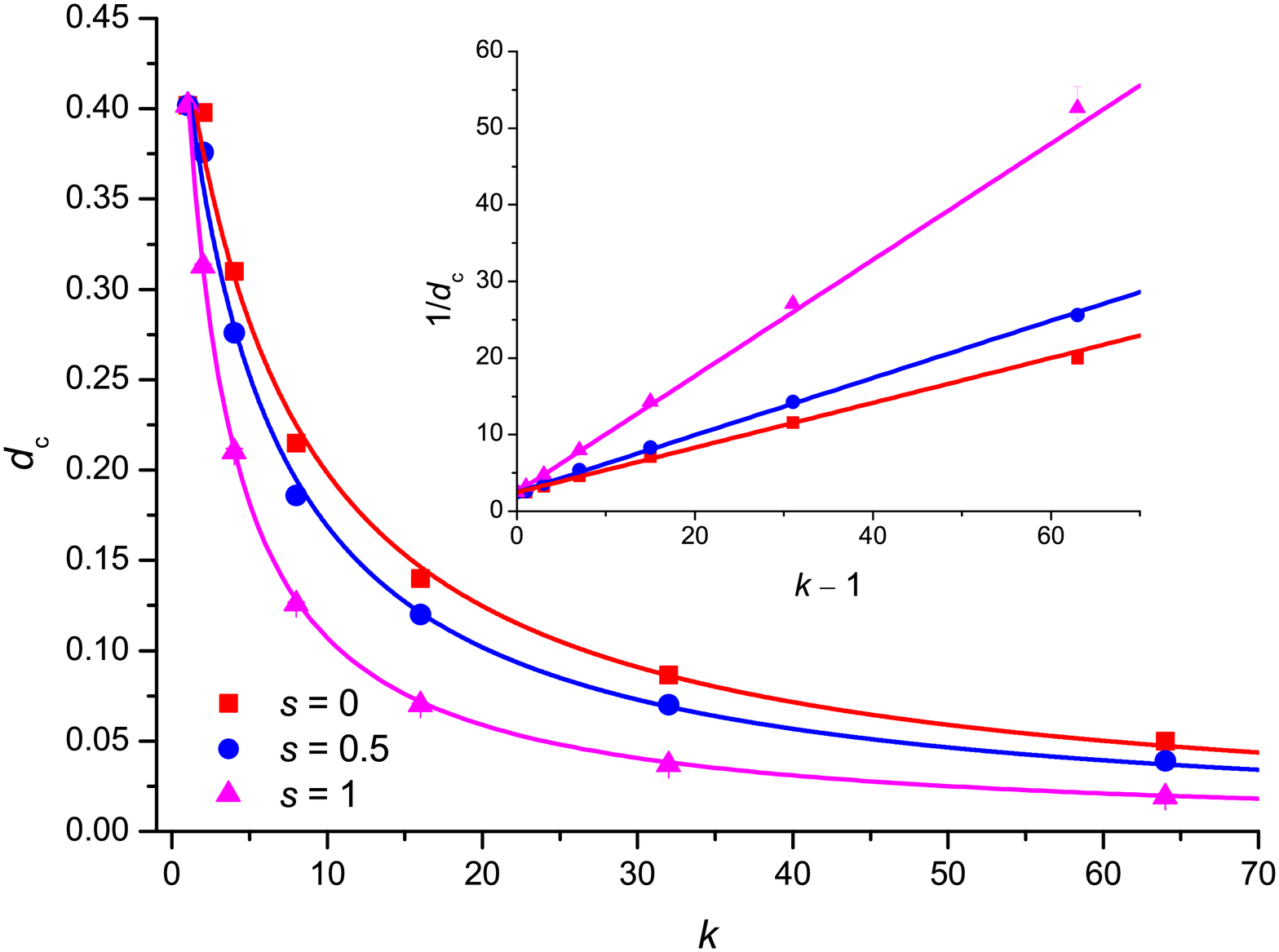}\\
\caption{Critical concentration of defects $d_c$ versus the value of $k$ for the RSC model at different values of the order parameter, $s$. Inset: $1/d_c$ versus $k$ dependencies.} \label{fig:f07}
\end{figure}

The critical concentration of defects, $d_c$, decreases as the values of $k$ and $s$ increase, and approaches  zero in the limit of very long $k$-mers, $k \to \infty$ (Fig.~\ref{fig:f07}). It is interesting that a rather good linear relation between $1/d_c$ and $k$ was observed at different values of $s$:
\begin{equation}\label{eq:linear}
d_c^{-1} = d_{c}^{-1}(k=1)+a(k-1),
\end{equation}
where $d_{c}^{-1}(k=1)=2.488\pm 0.002$ and $a=0.282\pm0.004$ ($s=0.0$), $a=0.371\pm0.004$ ($s=0.5$) and $a=0.794\pm0.001$ ($s=1.0$). Note, that the value $1-d_{c}(k=1)=0.598\pm 0.003$ is close to the value of the threshold concentration $\simeq 0.5927$ for ordinary monomer percolation~\cite{Stauffer}.
\begin{figure}[htbp]
  \centering
\includegraphics[width=0.9\linewidth]{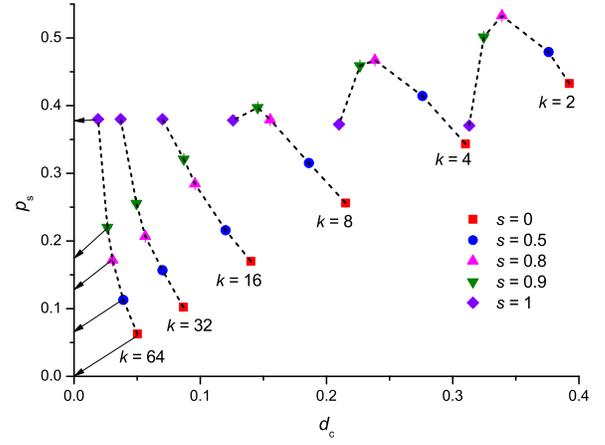}\\
\caption{Density of clusters, $p_s$, versus the critical concentration of defects, $d_c$, for the RSC model at different values of the order parameter, $s$, and of $k$. Arrows show the values of $p_s$ in the limit of very long $k$-mers, $k \to \infty$.} \label{fig:f08}
\end{figure}

The density of the clusters, $p_s$, at the critical concentration of defects, $d_c$, for the RSC model depends in a complex manner on $s$ and $k$ (Fig.~\ref{fig:f08}). For disordered systems ($s=0$), the value of $p_s$ tends to zero in the limits of the very long $k$-mers, $k \to \infty$ and very small critical concentrations $d_c \to 0$. The formation of such `empty' or loose clusters can be explained by analyzing the cluster patterns presented in Fig.~\ref{fig:f01}. In zero defect systems ($d=0$),  compact clusters form with significant stacks of identically oriented $k$-mers (horizontal and vertical). These stacks are responsible for the high values of $p_s$. The introduction of defects strongly prevents the formation of stacks (see, e.g., Fig.~\ref{fig:f01}a) and loose networks with near zero density are formed. On the other hand,  denser clusters are formed for ordered systems with $p_s\approx 0.065$ at $s=0.5$ and $p_s\approx 0.38$ at $s=1.0$.

\clearpage

\section{Final remarks and conclusions}\label{sec:discussion}
For the problem of $k$-mers deposition, the RRSA and the RSC models have huge differences in their deposition rules and structure of the clusters they generate. In the RRSA model, random deposition of $k$-mers is performed, with multiple clusters being formed that can consolidate in the course of their growth. Intensive studies with the RRSA have shown that the jamming concentration continuously decreases as the lengths of the $k$-mers  increase~\cite{Lebovka2011PRE}. The percolation threshold initially decreases and then increases with increasing values of $k$~\cite{Tarasevich2012PRE}. For a completely disordered system, i.e. at $s=0$, a conjecture has been offered that the formation of an infinite cluster is impossible when $k$ exceeds approximately  $1.2 \times 10^4$~\cite{Tarasevich2012PRE}.
On the other hand, in the RSC model the single cluster grow at its perimeter by an Eden growth process~\cite{Eden1961} and the formation of an infinite cluster is not restricted for any values of $k$. This model assumes strong attractions of the newcomer to the previously deposited particles at their lateral boundaries. The Eden-like clusters display noticeable shape anisotropy at high values of the order parameter, $s$.

It is amazing that, in spite of these differences between the RRSA and the RSC models, they display some general similarities. In the absence of defects on the substrate (for zero defect systems, $d=0$), the jamming concentration, $p_j$, for the RRSA model and density of the cluster, $p_s$, for the RSC model, display similar dependencies on the linear segment length, $k$, and on the order parameters (see, Fig.~\ref{fig:f06}). The impacts of defects on the percolation and jamming characteristics are also fairly similar. For both models the suppression of the growth of an infinite (percolation) cluster at some critical concentration of defects, $d_c$, can be observed. Phase diagrams in the form of the $p_c(d_c)$ (RRSA model) and $p_s(d_c)$ dependencies for different values of $s$ and $k$ were determined.

For the RRSA model, the value of $p_c$ ranges in the interval $\approx 0.465-0.58$ ($k=2-32$) and the value of $d_c$ decreases for short $k$-mers ($k < 16$) but increases for long $k$-mers ($k > 16$) as the value of $s$ increases. Moreover, for short $k$-mers, the percolation threshold is almost insensitive to the defect concentration for any values of $s$. For the RSC model the value of $p_s$ at the critical concentration of defects, $d_c$, depends in a complex manner on the values of $s$ and $k$. For disordered systems ($s=0$), the value of $p_s$ tends to zero in the limits of very long $k$-mers, $k \to \infty$. This reflects a suppression of $k$-mer stacking by the defects that results in the formation of loose clusters with very low density.

\section*{Acknowledgments}

The reported study was partially supported by the Russian Foundation for Basic Research, Research project No. 14-02-90402\_Ukr\_a, the Ministry of Education and Science of the Russian Federation, Project
No.~643, and the National Academy of Sciences of Ukraine, Project No.~43-02-14(U).

%\clearpage
\bibliography{RSA,percolation}

\end{document}